\documentstyle[11pt,epsfig]{article}
\date{}
\begin{document}
\title{{\bf Classical and quantum dynamics of a perfect fluid scalar-metric cosmology}}
\author{Babak Vakili\thanks{%
email: b-vakili@iauc.ac.ir}\\\\
{\small {\it Department of Physics, Azad University of Chalous,}}\\{\small {\it P.O. Box 46615-397, Chalous, Iran}}}
\maketitle
\hspace{1.3cm}{\it This work is dedicated to the revered memory of Prof.  Masoud Alimohammadi}
\begin{abstract}
We study the classical and quantum models of a Friedmann-Robertson-Walker (FRW) cosmology, coupled to a perfect
fluid, in the context of the scalar-metric gravity. Using the Schutz' representation for the perfect fluid, we show that, under a particular gauge choice, it may lead to the identification of a time parameter for the corresponding dynamical system. It is shown that the evolution of the universe based on the classical cosmology represents a late time power law expansion coming from a big-bang singularity in which the scale factor goes to zero while the scalar field blows up. Moreover, this formalism gives rise to a Schr\"{o}dinger-Wheeler-DeWitt (SWD)
equation for the quantum-mechanical description of the model under consideration, the eigenfunctions of which can be used to construct the wave function  of the universe. We use the resulting wave function  in order to investigate the possibility of the avoidance of classical singularities due to quantum effects by means of the many-worlds and ontological interpretation of quantum cosmology.\vspace{5mm}\newline PACS numbers: 04.20.Fy, 04.60.Kz, 98.80.Qc \vspace{0.8mm}\newline Keywords: quantum scalar cosmology, perfect fluid, Schutz formalism
\end{abstract}
\section{Introduction}
Classical and semiclassical scalar fields play an essential role in unified theories of interactions and also in all branches of the modern cosmological theories \cite{1}. From a cosmological point of view, there is a renewed interest in the scalar-tensor models in which a non-minimal coupling appears between the geometry of space-time and a scalar field \cite{2}. This is because that a number of different scenarios in cosmology such as spatially flat and accelerated expanding universe at the present time \cite{3}, inflation \cite{4}, dark matter and dark energy \cite{5}, and a rich variety of behaviors can be accommodated phenomenologically by scalar fields. Traditionally cosmological models of inflation use a single scalar field with a canonical kinetic term of the form $1/2 g^{\mu \nu}\partial_{\mu}\phi \partial_{\nu}\phi$ with some particular
self-interaction potential $V(\phi)$ like $1/2 m^2 \phi^2$ or $\lambda \phi^4$, etc. Such a scalar field is often known as minimally coupled to the geometry. On the other hand, the scalar field in what is qualified to be called the scalar-tensor theory is not simply added to the tensor
gravitational field, but comes into play through the non-minimal coupling term \cite{6}.

In this paper we shall study the classical and quantum time evolution of a flat FRW model with a perfect fluid matter source and a non-linear self-coupling scalar field minimally coupled to gravity. In our model the the scalar field is coupled to the metric through a coupling function $F(\phi)$ in order to obtain a scalar-metric theory which as we shall see leads to satisfactory cosmological results. The classical cosmology of such a formulation is studied in \cite{7} and it is shown that such scalar-metric coupling gives a satisfactory description of the weak fields and a possible way to remove the missing matter problem in non-flat cosmologies. Also, a generalized version of the model studied in  \cite{7} in classical and quantum cosmology level is considered in \cite{8}.

Here, we first consider a gravitational action in the FRW background
in which a scalar field is coupled to the metric with a generic form
of a $F(\phi)$ function. To make the model simple and solvable,
after some steps, we take a polynomial coupling of the form
$F(\phi)=\lambda \phi^m$. For the matter source of gravity, we
consider a perfect fluid in Schutz' formalism \cite{9}. The
advantage of using this formalism in our quantum cosmological model
is that, in a natural way, it can offer a time parameter in terms of
dynamical variables of the perfect fluid \cite{10}. Indeed, as we
shall show, after a canonical transformation the conjugate momentum
associated to one of the variables of the fluid appears linearly in
the Hamiltonian of the model. Therefore, canonical quantization
results in a Schr\"{o}dinger-Wheeler-DeWitt (SWD) equation, in which
this matter variable plays the role of time. In terms of this time
parameter, we shall obtain the dynamical behavior of the cosmic
scale factor and the scalar field. We show that the evolution of the
scale factor represents a late time power law expansion coming from
a big-bang singularity. Also, the classical behavior of the scalar
field shows a blow up in this regime. Finally, we deal with the
quantization of the model, and by computing the expectation values
of the scale factor and the scalar field and also their ontological
counterparts, we show that the evolution of the universe according
to the quantum picture is free of classical singularities.
\section{The classical model}
In this section we consider a FRW cosmology in which a scalar field
which is coupled to the metric. Also, a perfect fluid with which the
action of the model is augmented, plays the role of the matter part
of the model. In the context of the ADM formalism the action reads
(in what follows we work in units where $c=\hbar=16\pi G=1$)
\begin{equation}\label{A}
{\cal S}=\int_M d^4x \sqrt{-g}\left[R-F(\phi)g^{\mu \nu}\phi_{,\mu}\phi_{,\nu}\right]+2\int_{\partial M}d^3x
\sqrt{h}h_{ab}K^{ab}+\int_M d^4 x \sqrt{-g}p,\end{equation}where
$R$ is the scalar curvature and $F(\phi)$ is an arbitrary function of
the scalar field $\phi$. Also, $K^{ab}$ is the extrinsic curvature and $h_{ab}$ is the
induced metric over the three-dimensional spatial hypersurface,
which is the boundary $\partial M$ of the four-dimensional manifold
$M$. The last term of (\ref{A}) denotes the matter contribution to
the total action where $p$ is the pressure of perfect fluid which is
linked to its energy density by the equation of state
\begin{equation}\label{B}
p=\alpha \rho.\end{equation}
In Schutz' formalism the fluid's four-velocity is expressed in terms of five potentials $\epsilon$, $\zeta$, $\beta$, $\theta$ and $S$ as \cite{9}
\begin{equation}\label{C}
U_{\nu}=\frac{1}{\mu}(\epsilon_{,\nu}+\zeta \beta_{,\nu}+\theta
S_{,\nu}),\end{equation} where $\mu$ is the specific enthalpy, the
variable $S$ is the specific entropy while the potentials $\zeta$
and $\beta$ are related to torsion and are absent in the FRW models.
The variables $\epsilon$ and $\theta$ have no clear physical interpretation
in this formalism. The four-velocity satisfies the condition
\begin{equation}\label{D}
U^{\nu}U_{\nu}=-1.\end{equation}
We assume that the geometry of space–time is described by the FRW metric
\begin{equation}\label{E}
ds^2=-N^2(t)dt^2+a^2(t)\left[\frac{dr^2}{1-kr^2}+r^2(d\vartheta^2+
\sin^2 \vartheta d\varphi^2)\right],
\end{equation}
where $N(t)$ is the lapse function, $a(t)$ the scale factor and
$k$=$1$, $0$ and $-1$ correspond to the closed, flat and open
universe respectively. To proceed further, we need an effective
Lagrangian for the model whose variation with respect to its
dynamical variables yields the appropriate equations of motion.
Therefore, by considering the above action as representing a
dynamical system in which the scale factor $a$, scalar field $\phi$
and fluid's potentials play the role of independent dynamical
variables, we can rewrite the gravitational part of the action (\ref{A})
as
\begin{eqnarray}\label{F}
{\cal S}_{grav}=\int dt {\cal L}_{grav}(a,\dot{a},\phi,\dot{\phi})=
\int dt \left\{Na^3\left(R+\frac{1}{N^2}F(\phi)\dot{\phi}^2\right)-\lambda\left[R-\frac{6}{N^2}\left(\frac{\ddot{a}}{a}+
\frac{\dot{a}^2}{a^2}+\frac{k}{a^2}-\frac{\dot{N}\dot{a}}{Na}\right)\right]\right\},
\end{eqnarray}where we have introduced the definition of $R$ in terms of $a$ and its
derivatives as a constraint. This procedure allows us to remove the
second order derivatives from action (\ref{F}). The Lagrange
multiplier $\lambda$ can be obtained by variation with respect to
$R$, that is, $\lambda = Na^3$. Thus, we obtain the following
point-like Lagrangian for the gravitational part of the model
\begin{equation}\label{G}
{\cal
L}_{grav}=-\frac{6}{N}a\dot{a}^2+6kNa+\frac{1}{N}F(\phi)a^3\dot{\phi}^2.\end{equation}Also,
the matter part of the action (\ref{A}) becomes ${\cal
S}_{matt}=\int d^3xdt Na^3p$, so the Lagrangian density of the fluid
is ${\cal L}_{matt}=Na^3p$. Following the thermodynamic description
of \cite{9}, the basic thermodynamic relations take the form
\begin{equation}\label{baby1}
\rho=\rho_0(1+\Pi),\hspace{0.5cm}\mu=1+\Pi+\frac{p}{\rho_0},\end{equation}where
$\rho_0$ and $\Pi$ are the rest-mass density and the specific
internal energy of the fluid respectively. These quantities together
with the temperature of the system $\tau$, obey the first law of the
thermodynamics $\tau dS=d\Pi+pd(1/\rho_0)$, which can be rewritten
as
\begin{equation}\label{baby2}
\tau dS=d\Pi+pd\left(\frac{1}{\rho_0}\right)=(1+\Pi)d\left[\ln
\rho_0^{-\alpha}(1+\Pi)\right],\end{equation}where we have used the
equation of state (\ref{B}). Therefore, we obtain the following
expressions for the temperature and the entropy of the fluid
\begin{equation}\label{baby3}
\tau=1+\Pi,\hspace{0.5cm}S=\ln \rho_0^{-\alpha}(1+\Pi)=\ln
\frac{\mu}{\alpha+1}\rho_0^{-\alpha}.\end{equation}Now, we can
express $\rho_0$ and $\Pi$ as functions of $\mu$ and $S$ as
\begin{equation}\label{baby4}
1+\Pi=\frac{\mu}{\alpha+1},\hspace{0.5cm}\rho_0=\left(\frac{\mu}{\alpha+1}\right)^{1/\alpha}e^{-S/\alpha},\end{equation}so
that with the help of (\ref{baby1}), one can put the equation of
state in the form
\begin{equation}\label{baby5}
p=\frac{\alpha}{(\alpha+1)^{1+1/\alpha}}\mu^{1+1/\alpha}e^{-S/\alpha}.\end{equation}On
the other hand, normalization of the fluid's four-velocity
(\ref{C}), according to the relation (\ref{D}) implies
$\mu=(\dot{\epsilon}+\theta \dot{S})/N$. Therefore, using the above
constraints and  thermodynamical considerations for the fluid we
find
\begin{equation}\label{H}
{\cal
L}_{matt}=N^{-1/\alpha}a^3\frac{\alpha}{(\alpha+1)^{1+1/\alpha}}\left(\dot{\epsilon}+\theta
\dot{S}\right)^{1+1/\alpha}e^{-S/\alpha}.\end{equation}Let us now
construct the Hamiltonian for our model. The momenta conjugate to
each of the above variables can be obtained from the definition
$P_{q}=\frac{\partial {\cal L}}{\partial \dot{q}}$. In terms of the
conjugate momenta the Hamiltonian is given by
\begin{equation}\label{I}
H=H_{grav}+H_{matt}=\dot{a}P_a+\dot{\phi}P_{\phi}+\dot{\epsilon}P_{\epsilon}+\dot{S}P_S-{\cal
L},\end{equation}where ${\cal L}={\cal L}_{grav}+{\cal L}_{matt}$.
Expression (\ref{I}) leads to
\begin{equation}\label{J}
H=N{\cal
H}=N\left[-\frac{1}{24}\frac{P_a^2}{a}+\frac{1}{4F(\phi)a^3}P_{\phi}^2-6ka+a^{-3\alpha}e^SP_{\epsilon}^{\alpha+1}\right].\end{equation}
Now, consider the following canonical transformation which is a
generalization of the ones used in \cite{11}
\begin{equation}\label{K}
\begin{array}{cc}
T=-P_Se^{-S}P_{\epsilon}^{-(\alpha+1)}, & P_T=P_{\epsilon}^{\alpha+1}e^S, \\\\
\bar{\epsilon}=\epsilon-(\alpha+1)\frac{P_S}{P_{\epsilon}}, & \bar{P_{\epsilon}}=P_{\epsilon}.
\end{array}
\end{equation}Under this transformation Hamiltonian (\ref{J}) takes the form
\begin{equation}\label{L}
H=N{\cal H}=N\left[-\frac{1}{24}\frac{P_a^2}{a}+\frac{1}{4F(\phi)a^3}P_{\phi}^2-6ka+\frac{P_T}{a^{3\alpha}}\right].\end{equation}We
see that the momentum $P_T$ is the only remaining canonical variable
associated with the matter and appears linearly in the Hamiltonian. The
setup for constructing the phase space and writing the Lagrangian
and Hamiltonian of the model is now complete.

The classical dynamics is governed by the Hamiltonian equations, that is
\begin{eqnarray}\label{M}
\left\{
\begin{array}{ll}
\dot{a}=\{a,H\}=-\frac{N}{12}\frac{P_a}{a},\\\\
\dot{P_a}=\{P_a,H\}=N\left[-\frac{1}{24}\frac{P_a^2}{a^2}+\frac{3}{4F(\phi)a^4}P_{\phi}^2+6k +3\alpha a^{-3\alpha-1}P_T\right],\\\\
\dot{\phi}=\{\phi,H\}=\frac{N}{2F(\phi)a^3}P_{\phi},\\\\
\dot{P_{\phi}}=\{P_{\phi},H\}=N\frac{P_{\phi}^2}{4a^3}\frac{F'(\phi)}{F(\phi)^2},\\\\
\dot{T}=\{T,H\}=\frac{N}{a^{3\alpha}},\\\\
\dot{P_T}=\{P_T,H\}=0.
\end{array}
\right.
\end{eqnarray}
We also have the constraint equation ${\cal H}=0$. Up to this point
the cosmological model, in view of the concerning issue of time, has
been of course under-determined. Before trying to solve these
equations we must decide on a choice of time in the theory. The
under-determinacy problem at the classical level may be resolved by
using the gauge freedom via fixing the gauge. A glance at the above
equations shows that choosing the gauge $N=a^{3\alpha}$, we have
\begin{equation}\label{N}
N=a^{3\alpha}\Rightarrow T=t,\end{equation}which means that variable
$T$ may play the role of time in the model. Therefore, the classical
equations of motion can be rewritten in the gauge $N=a^{3\alpha}$ as
follows
\begin{eqnarray}\label{O}
\left\{
\begin{array}{ll}
\dot{a}=-\frac{1}{12}a^{3\alpha-1}P_a,\\\\
\dot{P_a}=-\frac{1}{24}a^{3\alpha-2}P_a^2+\frac{3}{4F(\phi)}a^{3\alpha-4}P_{\phi}^2+6 k a^{3\alpha}+3\alpha P_0a^{-1},\\\\
\dot{\phi}=\frac{P_{\phi}}{2F(\phi)}a^{3\alpha-3},\\\\
\dot{P_{\phi}}=\frac{1}{4}\frac{F'(\phi)}{F(\phi)^2}P_{\phi}^2 a^{3\alpha-3},
\end{array}
\right.
\end{eqnarray}where we take $P_T=P_0=\mbox{const.}$ from the last equation of (\ref{M}). The two last equations of the above system indicate that the field $\phi$ obeys the second order equation of motion
\begin{equation}\label{P}
\frac{\ddot{\phi}}{\dot{\phi}}+\frac{1}{2}\frac{F'(\phi)}{F(\phi)}\dot{\phi}+3(1-\alpha)\frac{\dot{a}}{a}=0,\end{equation}which is equivalent to the conservation law, provided the standard perfect fluid energy-momentum tensor is introduced. Equation (\ref{P}) can easily be integrated to yield
\begin{equation}\label{R}
\dot{\phi}^2 F(\phi)=C a^{6(\alpha-1)},\end{equation}where $C$ is an
integration constant. Also, eliminating the momenta from the system
(\ref{O}) results
\begin{equation}\label{S}
-6a^{1-3\alpha}\dot{a}^2+\dot{\phi}^2F(\phi)a^{3-3\alpha}-6ka^{1+3\alpha}+P_0=0,\end{equation}which
is nothing but the constraint equation ${\cal H}=0$. With the help
of (\ref{R}) this equation can be put into the form
\begin{equation}\label{T}
6\dot{a}^2=C a^{6\alpha-4}-6ka^{6\alpha}+P_0 a^{3\alpha-1},\end{equation}where for the flat case $k=0$, its solution reads
\begin{equation}\label{U}
a(t)=\left[\frac{3P_0(1-\alpha)^2}{8}(t-\delta)^2-\frac{C}{P_0}\right]^{\frac{1}{3(1-\alpha)}},\end{equation}for $\alpha \neq 1$, and
\begin{equation}\label{V}
a(t)=a_0 \exp\left(\sqrt{\frac{C+P_0}{6}}t\right),\end{equation}for
$\alpha=1$, with constants $\delta$ and $a_0$. We may set
$\delta=\frac{\sqrt{8C}}{\sqrt{3}P_0(1-\alpha)}$, so that
$a(t=0)=0$.  What remains to be found is an expression for the
scalar field $\phi(t)$. In the following, we shall consider the case
of a coupling function in the form $F(\phi)=\lambda \phi^m$. With
this choice for the function $F(\phi)$ we are able to calculate the
time evolution of the scalar field as
\begin{equation}\label{W}
\phi(t)=\left\{
\begin{array}{ll}
\left\{\frac{\sqrt{2}(m+2)}{\sqrt{3\lambda}(1-\alpha)}\tanh^{-1}\left[\frac{\sqrt{3}P_0(1-\alpha)}{\sqrt{8C}}(t-\delta)\right]\right\}^{\frac{2}{m+2}}
,\hspace{0.5cm}m\neq -2,\\\\
\exp\left\{\frac{2\sqrt{2}}{\sqrt{3\lambda}(1-\alpha)}\tanh^{-1}\left[\frac{\sqrt{3}P_0(1-\alpha)}{\sqrt{8C}}(t-\delta)\right]\right\},\hspace{0.5cm}m=-2,
\end{array}
\right.
\end{equation}for $\alpha \neq 1$ and
\begin{equation}\label{X}
\phi(t)=\left\{
\begin{array}{ll}
\left[\sqrt{\frac{C}{\lambda}}\frac{m+2}{2}(t-\delta)\right]^{\frac{2}{m+2}},\hspace{0.5cm}m\neq -2,\\\\
\phi_0e^{\sqrt{\frac{C}{\lambda}}(t-\delta)},\hspace{0.5cm}m=-2,
\end{array}
\right.
\end{equation}for $\alpha =1$.
We see that for $\alpha \neq 1$, the evolution of the universe based
on (\ref{U}) begins with a big-bang singularity at $t=0$ and, for
$\alpha <1$, follows the power law expansion $a(t)\sim
t^{\frac{2}{3(1-\alpha)}}$ at late time of cosmic evolution while
the scalar field has a monotonically increasing behavior coming from
$\phi\rightarrow -\infty$, reaches zero and then blows up at a
finite time. Also, in the case of $\alpha=1$ (stiff matter), an
exponential solution is obtained for the corresponding cosmology in
the chosen time variable. In terms of cosmic time $\eta=\int N dt$,
solution (\ref{V}) reads as $a(\eta)\sim \eta^{1/3}$ which shows a
decelerated expansion. This not surprising since for $\alpha=1$
there is no violation of the strong energy condition, and hence an
accelerated expansion cannot be obtained. To understand the
relation between the big-bang singularity $a\rightarrow 0$ and the
blow up singularity $\phi\rightarrow \pm \infty$, we are going to
find a classical trajectory in configuration space $(a,\phi)$, where
the time parameter $t$ is eliminated. From (\ref{R}) and (\ref{T})
one has
\begin{equation}\label{AA}
\frac{\sqrt{F(\phi)}d\phi}{da}=\pm
\frac{\sqrt{C}a^{3(\alpha-1)}}{\sqrt{\frac{C}{6}a^{6\alpha-4}-ka^{6\alpha}+\frac{P_0}{6}a^{3\alpha-1}}},\end{equation}
where for the case $k=0$, after integration  reads
\begin{equation}\label{BB}
\phi(a)=\left\{\pm
\frac{\sqrt{2}(m+2)}{\sqrt{3\lambda}(1-\alpha)}\sinh^{-1}\left[\sqrt{\frac{C}{P_0}}a^{\frac{3(\alpha-1)}{2}}\right]\right\}^{\frac{2}{m+2}}.\end{equation}
Equation (\ref{BB}) describes two branches for which $\phi
\rightarrow 0$ (or a non-zero constant if we add a non-zero
integration constant to the above relation) if $a\rightarrow \infty$
and $\phi \rightarrow \pm \infty$ if $a\rightarrow 0$. This means
that at the late time limit, the scalar field approaches a constant
value while its blow up behavior corresponds to the big-bang
singularity. We shall see in the next section that this classical
picture will be modified if one takes into account the quantum
mechanical considerations in the problem at hand.
\section{The quantum model}
We now focus attention on the study of the quantum cosmology of the
model described above. We start by writing the Wheeler-DeWitt
equation from Hamiltonian (\ref{L}). A remark about our quantization procedure is that the canonical
transformation (\ref{K}) is applied to the classical Hamiltonian
(\ref{J}), resulting in Hamiltonian (\ref{L}) which we are going to
quantize. To make this acceptable, one should show that in the
quantum theory the two Hamiltonians are connected by some unitary
transformation, i.e. the transformation (\ref{K}) is also a quantum
canonical transformation. A quantum canonical transformation is
defined as a change of the phase space variables $(q,p)\rightarrow
(q',p')$ which preserves the Dirac bracket \cite{12}
\begin{equation}\label{Y}
\left[q,p\right]=i=\left[q'(q,p),p'(q,p)\right].\end{equation}Such a
transformation is implemented by a function $C(q, p)$ such
that
\begin{equation}\label{Z}
q'(q,p)=CqC^{-1},\hspace{0.5cm}p'(q,p)=CpC^{-1}.\end{equation}This
canonical transformation $C$, transforms the Hamiltonian as
$H'(q,p)=CH(q,p)C^{-1}$. For our case the canonical relations
$[S,P_S]=[\epsilon,P_{\epsilon}]=i$
yield
\begin{equation}\label{AB}
\left[T,P_T\right]=\left[-P_Se^{-S}P_{\epsilon}^{-(\alpha+1)},P_{\epsilon}^{\alpha+1}e^S\right]=-\left[P_Se^{-S},e^S\right]=
\left[e^S,P_S\right]e^{-S}=ie^Se^{-S}=i,\end{equation} and
\begin{equation}\label{AC}
\left[\bar{\epsilon},\bar{P_{\epsilon}}\right]=
\left[\epsilon-(\alpha+1)P_SP_{\epsilon}^{-1},P_{\epsilon}\right]=\left[\epsilon,P_{\epsilon}\right]=i,\end{equation}which
means that the transformation (\ref{K}) preserves the Dirac brackets
and thus is a quantum canonical transformation. Therefore, use of
the transformed Hamiltonian (\ref{L}) for quantization of the model
is quite reasonable.
\subsection{Schr\"{o}dinger-Wheeler-DeWitt equation}
\begin{figure}
\begin{tabular}{c}\hspace{-1cm}\epsfig{figure=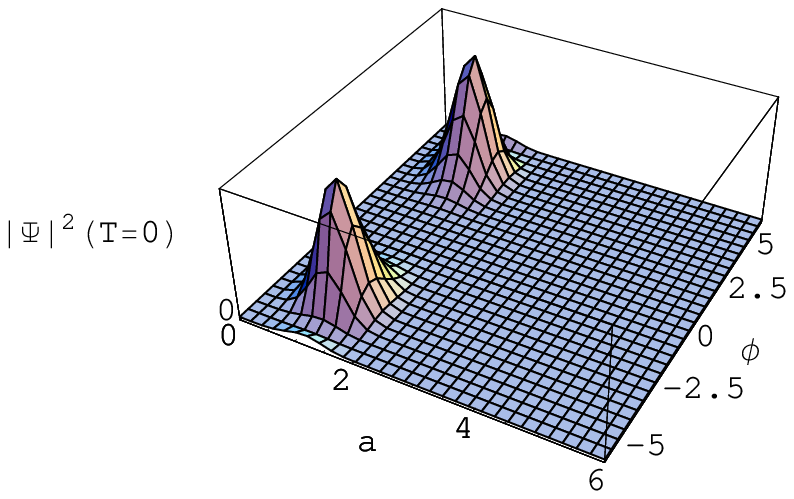,width=7cm}
\hspace{1cm} \epsfig{figure=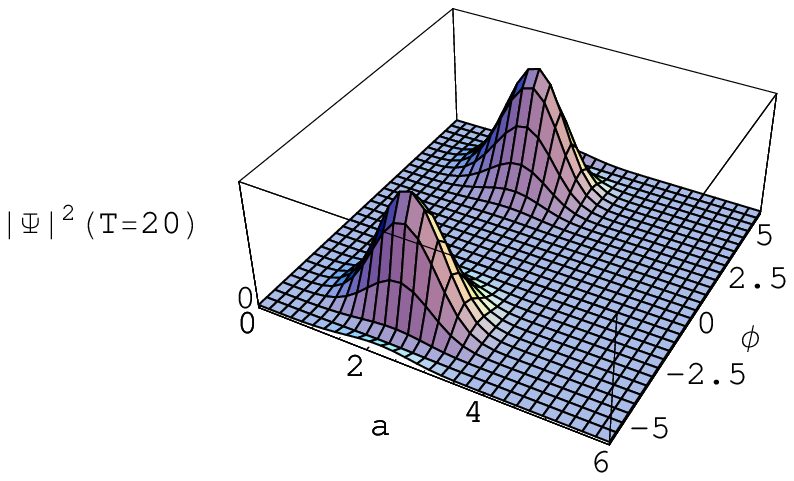,width=7cm}
\end{tabular}
\begin{tabular}{c}\hspace{-1cm}\epsfig{figure=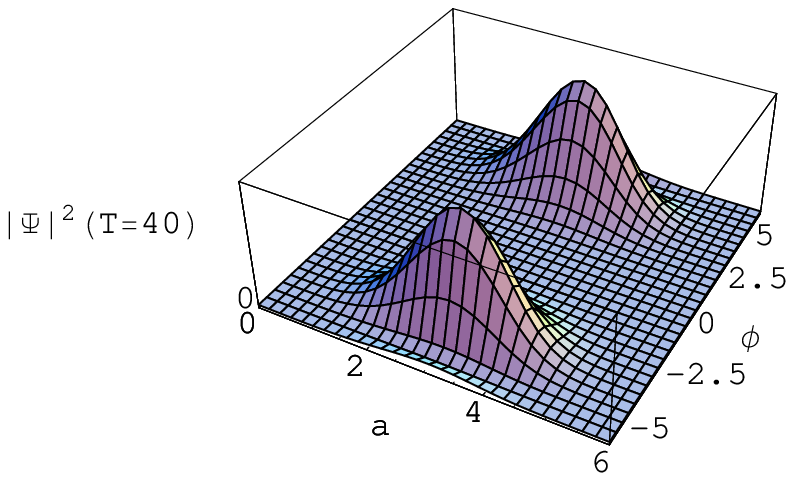,width=7cm}
\hspace{1cm} \epsfig{figure=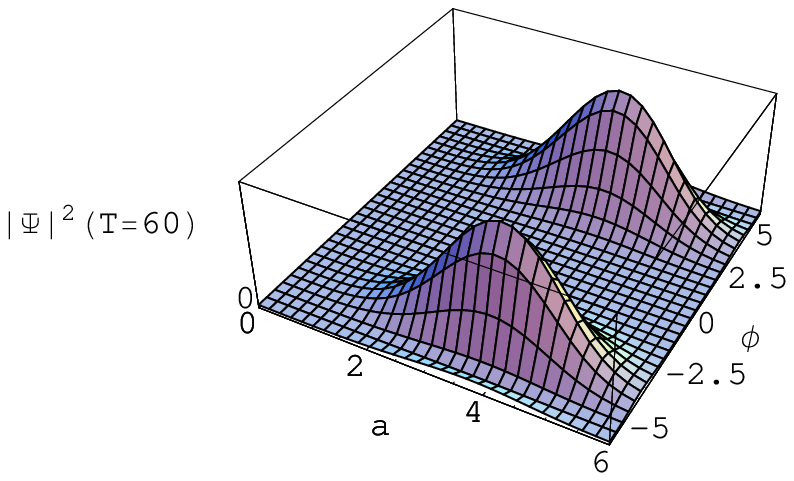,width=7cm}
\end{tabular}
\caption{\footnotesize  The figures show
$|\Psi(a,\phi,T)|^2$, the square of the wave function  in four different time parameter $T=0,20,40,60$.
The figures are plotted for the numerical values $\gamma = 1$, $m = 2$, $s=1$, $\lambda=1$, and we set the equation of state parameter $\alpha=-1/3$. After examining some other values for this parameter, we verify that the general behavior of the wave function  is repeated.}\label{fig1}
\end{figure}

Since the lapse function $N$ appears as a Lagrange multiplier in the Hamiltonian (\ref{L}), we have the Hamiltonian
constraint ${\cal H}=0$. Thus, application of the Dirac quantization procedure demands that the quantum states of the universe should be annihilated by the operator version of ${\cal H}$, that is
\begin{equation}\label{AC}
{\cal H}\Psi(a,\phi,T)=\left[-\frac{1}{24}\frac{P_a^2}{a}+\frac{1}{4F(\phi)a^3}P_{\phi}^2-6ka+\frac{P_T}{a^{3\alpha}}\right]\Psi(a,\phi,T)=0,
\end{equation}where $\Psi(a,\phi,T)$ is the wave function  of the universe. Choice of the ordering $a^{-1}P_a^2=P_a a^{-1}P_a$ and $F(\phi)^{-1}P_{\phi}^2=P_{\phi}F(\phi)^{-1}P_{\phi}$ to make the Hamiltonian Hermitian and use of the usual representation $P_q \rightarrow -i\partial_q$ results in
\begin{equation}\label{AD}
\left[a^{-1}\frac{\partial^2}{\partial a^2}-a^{-2}\frac{\partial}{\partial a}-6a^{-3}F(\phi)^{-1}\frac{\partial^2}{\partial \phi^2}+6a^{-3}\frac{F'(\phi)}{F(\phi)^2}\frac{\partial}{\partial \phi}-24ia^{-3\alpha}\frac{\partial}{\partial T}\right]\Psi(a,\phi,T)=0.\end{equation}This equation takes the form of a Schr\"{o}dinger equation $i\partial \Psi/\partial T=H\Psi$, in which the Hamiltonian operator is Hermitian with the standard inner product
\begin{equation}\label{AE}
<\Phi\mid \Psi>=\int_{(a,\phi)} a^{-3\alpha}\Phi^{\ast}\Psi da d\phi.\end{equation}We separate the variables in the SWD equation (\ref{AD}) as
\begin{equation}\label{AF}
\Psi(a,\phi,T)=e^{iET}\psi(a,\phi),\end{equation}leading to
\begin{equation}\label{AG}
\left[a^2\frac{\partial^2}{\partial a^2}-a\frac{\partial}{\partial a}-\frac{6}{F(\phi)}\frac{\partial^2}{\partial \phi^2}+6\frac{F'(\phi)}{F(\phi)^2}\frac{\partial}{\partial \phi}-24Ea^{3-3\alpha}\right]\psi(a,\phi)=0,\end{equation}where $E$ is a separation constant. The solutions of the above differential equation are separable and may be written in the form $\psi(a,\phi)=U(a)V(\phi)$ which yields
\begin{eqnarray}\label{AH}
\left\{
\begin{array}{ll}
\left[a^2\frac{d^2}{da^2}-a\frac{d}{da}+(24Ea^{3-3\alpha}+\nu^2)\right]U(a)=0,\\\\
\left[\frac{6}{F(\phi)}\frac{d^2}{d\phi^2}-6\frac{F'(\phi)}{F(\phi)^2}\frac{d}{d\phi}+\nu^2\right]V(\phi)=0,
\end{array}
\right.
\end{eqnarray}where $\nu$ is another constant of separation. Upon substituting the relation $F(\phi)=\lambda \phi^m$ into the above system, its solutions read in terms of Bessel functions $J$ and $Y$ as
\begin{eqnarray}\label{AI}
\left\{
\begin{array}{ll}
U(a)=a\left[c_1J_{\frac{2\sqrt{1-\nu^2}}{3(1-\alpha)}}\left(\frac{\sqrt{96E}}{3(1-\alpha)}a^{\frac{3(1-\alpha)}{2}}\right)+
c_2Y_{\frac{2\sqrt{1-\nu^2}}{3(1-\alpha)}}\left(\frac{\sqrt{96E}}{3(1-\alpha)}a^{\frac{3(1-\alpha)}{2}}\right)\right],\\\\
V(\phi)=\phi^{\frac{m+1}{2}}\left[d_1J_{\frac{m+1}{m+2}}\left(\frac{\nu\sqrt{6\lambda}}{3(m+2)}\phi^{\frac{m+2}{2}}\right)+
d_2Y_{\frac{m+1}{m+2}}\left(\frac{\nu\sqrt{6\lambda}}{3(m+2)}\phi^{\frac{m+2}{2}}\right)\right],\end{array}
\right.
\end{eqnarray}where $c_i (i=1,2)$ and $d_i (i=1,2)$ are integration constants. Thus, the eigenfunctions of the SWD equation can be written as
\begin{equation}\label{AJ}
\Psi_{\nu E}(a,\phi,T)=e^{iET}a \phi^{\frac{m+1}{2}} J_{\frac{2\sqrt{1-\nu^2}}{3(1-\alpha)}}\left(\frac{\sqrt{96E}}{3(1-\alpha)}a^{\frac{3(1-\alpha)}{2}}\right)
J_{\frac{m+1}{m+2}}\left(\frac{\nu\sqrt{6\lambda}}{3(m+2)}\phi^{\frac{m+2}{2}}\right),
\end{equation}where we have chosen $c_2=d_2=0$ for having well-defined functions in all ranges of variables $a$ and $\phi$.
We may now write the general solutions to the SWD equations as a superposition of the eigenfunctions, that is
\begin{equation}\label{AL}
\Psi(x,y,T)=\int_{E=0}^{\infty}\int_{\nu=0}^1 A(E)C(\nu)\Psi_{\nu E}(x,y,T)dE d\nu,
\end{equation}
where $A(E)$ and $C(\nu)$ are suitable weight functions to construct
the wave packets. By using the equality \cite{13}
\begin{equation}\label{AM}
\int_0^\infty e^{-a r^2}r^{\nu+1}J_{\nu}(br)dr=\frac{b^{\nu}}{(2a)^{\nu+1}}e^{-\frac{b^2}{4a}},\end{equation}
we can evaluate the integral over $E$ in (\ref{AL}) and simple
analytical expression for this integral is found if we choose the function $A(E)$ to be
a quasi-Gaussian weight factor
\begin{equation}\label{AN}
A(E)=\frac{16}{3(1-\alpha)^2}\left(\frac{\sqrt{96E}}{3(1-\alpha)}\right)^{\sqrt{1-\nu^2}}\exp \left(-\frac{32\gamma}{3(1-\alpha)^2}E\right),\end{equation} which results in
\begin{eqnarray}\label{AO}
\int_0^{\infty}A(E)e^{iET}J_{\frac{2\sqrt{1-\nu^2}}{3(1-\alpha)}}\left(\frac{\sqrt{96E}}{3(1-\alpha)}a^{\frac{3(1-\alpha)}{2}}\right)dE&=& \nonumber
\\a^{\sqrt{1-\nu^2}}\left[2\gamma-i\frac{3}{16}(1-\alpha)^2T\right]^{-1-\frac{2\sqrt{1-\nu^2}}{3(1-\alpha)}}
\exp\left[-\frac{a^{3(1-\alpha)}}{4\gamma-i\frac{3}{8}(1-\alpha)^2T}\right],
\end{eqnarray}where $\gamma$ is an arbitrary positive constant. Substitution of the above relation into equation (\ref{AL}) leads to the following expression for the wave function
\begin{eqnarray}\label{AP}
\Psi(a,\phi,T)&=&a\left[2\gamma-i\frac{3}{16}(1-\alpha)^2T\right]^{-1}\phi^{\frac{m+1}{2}}
\exp\left[-\frac{a^{3(1-\alpha)}}{4\gamma-i\frac{3}{8}(1-\alpha)^2T}\right]\nonumber \\ && \times \int_0^1 C(\nu)a^{\sqrt{1-\nu^2}}\left[2\gamma-i\frac{3}{16}(1-\alpha)^2T\right]^{-\frac{2\sqrt{1-\nu^2}}{3(1-\alpha)}}
J_{\frac{m+1}{m+2}}\left(\frac{\nu\sqrt{6\lambda}}{3(m+2)}\phi^{\frac{m+2}{2}}\right)d\nu.
\end{eqnarray}
To achieve an analytical closed expression for the wave function, we
assume that the above superposition is taken over such values of
$\nu$ for which one can use the approximation $\sqrt{1-\nu^2} \sim
1$. Now, by using the equality \cite{13}
\begin{equation}\label{AQ}
\int_0^1 \nu^{r+1} (1-\nu^2)^{s/2} J_r(z\nu)d\nu=\frac{2^s \Gamma
(s+1)}{z^{s+1}}J_{r+s+1}(z),\end{equation}and choosing  the weight
function $C(\nu)=\nu^{\frac{2m+3}{m+2}}(1-\nu^2)^{s/2}$, we are led
to the following expression for the wave function  \footnote{One may
have some doubts on this final form for the wave function and the
following results due to the assumption $\sqrt{1-\nu^2} \sim 1$.
Indeed, since the wave function can be a complex function, we may
extend the integration domain over $\nu$ to $0\leq\nu<\infty$. In
this case with a numerical study of equation (\ref{AP}), we have
verified that the general patterns of the resulting wave packets
follow the behavior shown in figure \ref{fig1} with a very good
approximation.}
\begin{eqnarray}\label{AR}
\Psi(a,\phi,T)&=&{\cal N}a^2\left[2\gamma-i\frac{3}{16}(1-\alpha)^2T\right]^{\frac{3\alpha-5}{3(1-\alpha)}}\phi^{-\frac{s(m+2)+1}{2}}
\exp\left[-\frac{a^{3(1-\alpha)}}{4\gamma-i\frac{3}{8}(1-\alpha)^2T}\right] \nonumber \\ && \times
J_{\frac{2m+3}{m+2}+s}\left(\frac{\sqrt{6\lambda}}{3(m+2)}\phi^{\frac{m+2}{2}}\right),
\end{eqnarray}where $s$ and ${\cal N}$
are an arbitrary constant and a numerical factor respectively. Now,
having the above expression for the wave function  of the universe,
we are going to obtain the predictions for the behavior of the
dynamical variables in the corresponding cosmological model. To do
this, in the next subsection, we shall adopt two approaches to
evaluate the classical behavior of the dynamical variables in the
model which lead to the same results. In the many-worlds
interpretation of quantum mechanics \cite{14}, we can calculate the
expectation values of the dynamical variables and, in the realm of
the ontological interpretation of quantum mechanics \cite{15}, one
can evaluate the Bohmian trajectories for those variables.
In figure \ref{fig1} we have plotted the square of the
wave function  for typical numerical values of the parameters.
As this figure shows, at $T=0$, the wave function  has two dominant
peaks in the vicinity of some non-zero values of $a$ and $\phi$. This
means that the wave function  predicts the emergence of the universe
from a state corresponding to one of its dominant peaks. However,
the emergence of several peaks in the wave packet may be interpreted
as a representation of different quantum states that may communicate
with each other through tunneling. This means that there are
different possible universes (states) from which our present
universe could have evolved and tunneled in the past, from one
universe (state) to another. As time progresses, the wave packet
begins to propagate in the $a$-direction, its width becoming wider
and its peaks moving with a group velocity towards the greater
values of $a$ while the values of $\phi$ remaining almost constant.
The wave packet disperses as time passes, the minimum width being
attained at $T = 0$. As in the case of the free particle in quantum
mechanics, the more localized the initial state at $T = 0$, the more
rapidly the wave packet disperses. Therefore, the quantum effects
make themselves felt only for small enough $T$ corresponding to
small $a$, as expected and the wave function  predicts that the
universe will assume states with larger $a$ and an almost constant
$\phi$ in its late time evolution.

\subsection{Recovery of the classical solutions}
In general, one of the most important features in quantum cosmology
is the recovery of classical cosmology from the corresponding
quantum model or, in other words, how can the WD wave function s
predict a classical universe. In this approach, one usually
constructs a coherent wavepacket with good asymptotic behavior in
the minisuperspace, peaking in the vicinity of the classical
trajectory. On the other hand, in an another approach to show the
correlations between classical and quantum pattern, following the
many-worlds interpretation of quantum mechanics \cite{14}, one may
calculate the time dependence of the expectation value of a
dynamical variable $q$ as
\begin{equation}\label{AS}
<q>(t)=\frac{<\Psi|q|\Psi>}{<\Psi|\Psi>}.\end{equation}Following this approach, we may write the expectation value for the scale factor as
\begin{equation}\label{AT}
<a>(T)=\frac{\int_{a=0}^{\infty}\int_{\phi=-\infty}^{\infty}a^{-3\alpha}\Psi^*a\Psi da d\phi}{\int_{a=0}^{\infty}\int_{\phi=-\infty}^{\infty}a^{-3\alpha}\Psi^*\Psi da d\phi},\end{equation}
which yields
\begin{equation}\label{AU}
<a>(T)=\frac{\Gamma
(\frac{2-\alpha}{1-\alpha})}{\Gamma(\frac{5-3\alpha}{3-3\alpha})}\left[2\gamma
+\frac{9}{512\gamma}(1-\alpha)^4
T^2\right]^{\frac{1}{3(1-\alpha)}}.\end{equation} It is important to
classify the nature of the quantum model as concerns the presence or
absence of singularities. For the wave function  (\ref{AR}), the
expectation value (\ref{AU}) of $a$ never vanishes, showing that
these states are nonsingular. Indeed, the expression (\ref{AU}) for
$\alpha <1$, represents a bouncing universe with no singularity
where its late time behavior coincides to the late time behavior of
the classical solution (\ref{U}), that is $a(t)\sim
t^{\frac{2}{3(1-\alpha)}}$. Now we can calculate the dispersion of
the wave packet in the $a$-direction which is defined as
\begin{equation}\label{dis1}
(\Delta a)^2=<a^2>-<a>^2,\end{equation}using (\ref{AR}) and (\ref{AU}), we get
\begin{equation}\label{dis2}
(\Delta a)^2 \sim \left[2\gamma +\frac{9}{512\gamma}(1-\alpha)^4 T^2\right]^{\frac{2}{3(1-\alpha)}}.\end{equation}The result is that the wave packet traveling in the $a$-direction, spreads as time increases and thus its degree of localization is reduced. The width of the wave packet evaluated in (\ref{dis2}) agree with the discussion in the end of the previous subsection. Indeed, we may interpret the above relation for the width of the wave function  as the coincidence of the classical trajectories with the quantum ones for large values of time.
Also, the expectation value for the scalar field reads as
\begin{equation}\label{AV}
<\phi>(T)=\frac{\int_{a=0}^{\infty}\int_{\phi=-\infty}^{\infty}a^{-3\alpha}\Psi^*\phi\Psi da d\phi}{\int_{a=0}^{\infty}\int_{\phi=-\infty}^{\infty}a^{-3\alpha}\Psi^*\Psi da d\phi},\end{equation}with the result
\begin{equation}\label{AW}
<\phi>(T)=\frac{\int \phi^{-s(m+2)}\left[J_{\frac{2m+3}{m+2}}\left(\frac{\sqrt{6\lambda}}{3(m+2)}\phi^{\frac{m+2}{2}}\right)\right]^2d\phi}
{\int \phi^{-s(m+2)-1}\left[J_{\frac{2m+3}{m+2}}\left(\frac{\sqrt{6\lambda}}{3(m+2)}\phi^{\frac{m+2}{2}}\right)\right]^2d\phi}=\mbox{const}..
\end{equation}
We see that the expectation value of $\phi$ does not depend on time
which is just the behavior predicted by the wave function  of the
SWD equation. This result is comparable with those obtained in
\cite{14-1} where a constant expectation value for the dilatonic
field in a quantum cosmological model based on the string effective
action coupled to matter has been obtained. From the classical
solutions in the previous section, it is clear that this is the
classical trajectory obtained from (\ref{BB}) in the limit
$a\rightarrow \infty$. Therefore, in view of the behavior of the
scale factor and the scalar field, the classical solutions
(\ref{AU}) and (\ref{AW}) are in complete agreement with the quantum
patterns shown in figure \ref{fig1}, and both predict a
(nonsingular) monotonically increasing evolution for the scale
factor and consequently there is an almost good correlation between
the quantum patterns and classical trajectories.

The issue of the correlation between classical and quantum schemes
may be addressed from another point of view. It is known that the
results obtained by using the many-world interpretation agree with
those that can be obtained using the ontological interpretation of
quantum mechanics \cite{15}. In Bohmian interpretation, the wave
function  is written as
\begin{equation}\label{AX}
\Psi(a,\phi,T)=\Omega (a,\phi,T)e^{iS(a,\phi,T)},\end{equation}where $\Omega$ and $S$ are some real functions. Substitution of this expression into the SWD equation (\ref{AD}) leads to the continuity equation
\begin{equation}\label{AY}
2a^2\frac{\partial \Omega}{\partial a}\frac{\partial S}{\partial a}+a^2\Omega\frac{\partial^2 S}{\partial a^2}+a\Omega \frac{\partial S}{\partial a}-\frac{12}{F(\phi)}\frac{\partial \Omega}{\partial \phi}\frac{\partial S}{\partial \phi}-\frac{6}{F(\phi)}\Omega \frac{\partial^2 S}{\partial \phi^2}+6\frac{F'(\phi)}{F(\phi)^2}\Omega \frac{\partial S}{\partial \phi}-24a^{3(1-\alpha)}\frac{\partial \Omega}{\partial T}=0,
\end{equation}and the modified Hamilton-Jacobi equation
\begin{equation}\label{AZ}
-\frac{1}{24}\frac{1}{a}\left(\frac{\partial S}{\partial a}\right)^2+\frac{1}{4F(\phi)a^3}\left(\frac{\partial S}{\partial \phi}\right)^2+a^{-3\alpha}\left(\frac{\partial S}{\partial T}\right)+{\cal Q}=0,
\end{equation}in which the quantum potential ${\cal Q}$ is defined as
\begin{equation}\label{BA}
{\cal Q}=\frac{1}{24a\Omega}\frac{\partial^2 \Omega}{\partial a^2}+\frac{1}{24 a^2 \Omega}\frac{\partial \Omega}{\partial a}-\frac{1}{4a^3 F(\phi)\Omega}\frac{\partial^2 \Omega}{\partial \phi^2}+\frac{F'(\phi)}{4a^3F(\phi)^2\Omega}\frac{\partial \Omega}{\partial \phi}.
\end{equation} The real functions $\Omega (a,\phi,T)$ and $S(a,\phi,T)$ can be obtained from the wave function  (\ref{AR}) as
\begin{equation}\label{BC}
\Omega=\frac{a^2 \phi^{-\frac{s(m+2)+1}{2}}}
{\left[4\gamma^2+\frac{9}{256}(1-\alpha)^4T^2\right]^{\frac{5-3\alpha}{6(1-\alpha)}}}\exp\left[-\frac{4\gamma a^{3(1-\alpha)}}{16\gamma^2+\frac{9}{64}(1-\alpha)^4 T^2}\right]J_{\frac{2m+3}{m+2}+s}\left(\frac{\sqrt{6\lambda}}{3(m+2)}\phi^{\frac{m+2}{2}}\right),
\end{equation}
\begin{equation}\label{BD}
S=-\frac{3}{8}\frac{a^{3(1-\alpha)}(1-\alpha)^2 T}{16\gamma^2 +\frac{9}{64}(1-\alpha)^4 T^2}+\frac{5-3\alpha}{3(1-\alpha)}\arctan \left[\frac{3(1-\alpha)^2 T}{32 \gamma}\right].
\end{equation}In this interpretation the classical trajectories, which determine the behavior of the
scale factor and scalar field are given by
\begin{equation}\label{BE}
P_a=\frac{\partial S}{\partial a},\hspace{0.5cm}P_{\phi}=\frac{\partial S}{\partial \phi}.\end{equation}Using the expressions for $P_a$ and $P_{\phi}$ in (\ref{O}), the equations for the classical trajectories become
\begin{equation}\label{BF}
-12a^{1-3\alpha}\dot{a}=-\frac{9}{8}\frac{(1-\alpha)^3 T a^{2-3\alpha}}{16\gamma^2+\frac{9}{64}(1-\alpha)^4 T^2},
\end{equation}and
\begin{equation}\label{BG}
2F(\phi)a^{3(1-\alpha)}\dot{\phi}=0.\end{equation}Therefore, after integration we get
\begin{equation}\label{BH}
a(T)=a_0\left[16\gamma^2+\frac{9}{64}(1-\alpha)^4 T^2\right]^{\frac{1}{3(1-\alpha)}},\end{equation}and
\begin{equation}\label{BI}
\phi(T)=\mbox{const.},\end{equation}where $a_0$ is a constant of integration. These solutions have the same behavior as the expectation values computed in (\ref{AU}) and (\ref{AW}) and like those are free of singularity.
\begin{figure}
\begin{tabular}{c}\hspace{-1cm}\epsfig{figure=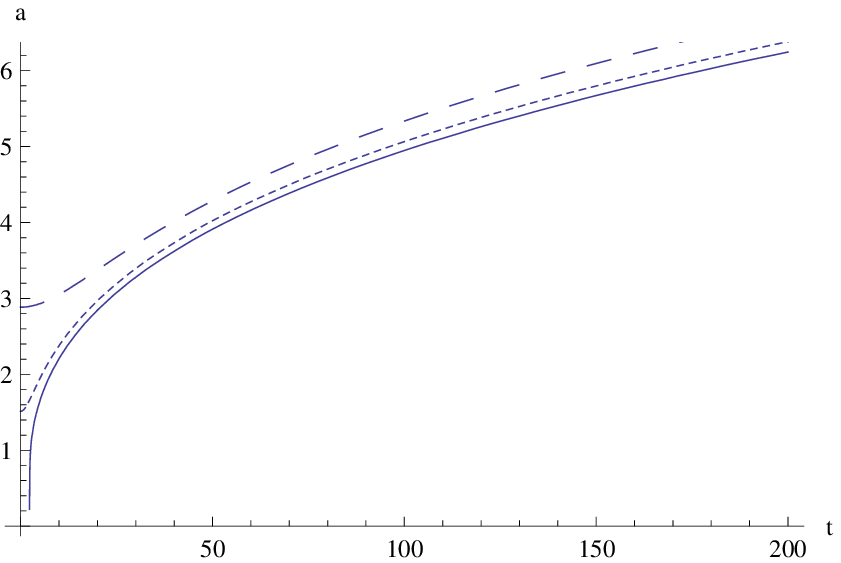,width=7cm}
\hspace{1cm} \epsfig{figure=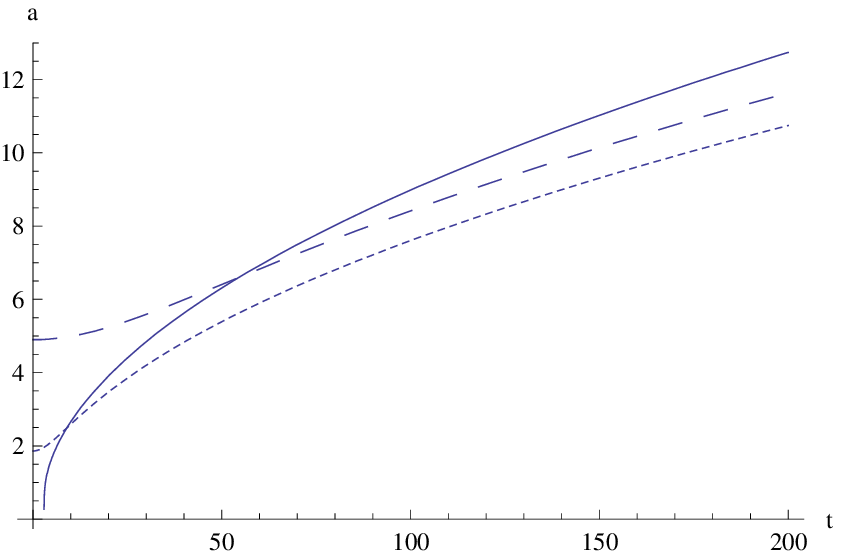,width=7cm}
\end{tabular}
\begin{tabular}{c}\hspace{-1cm}\epsfig{figure=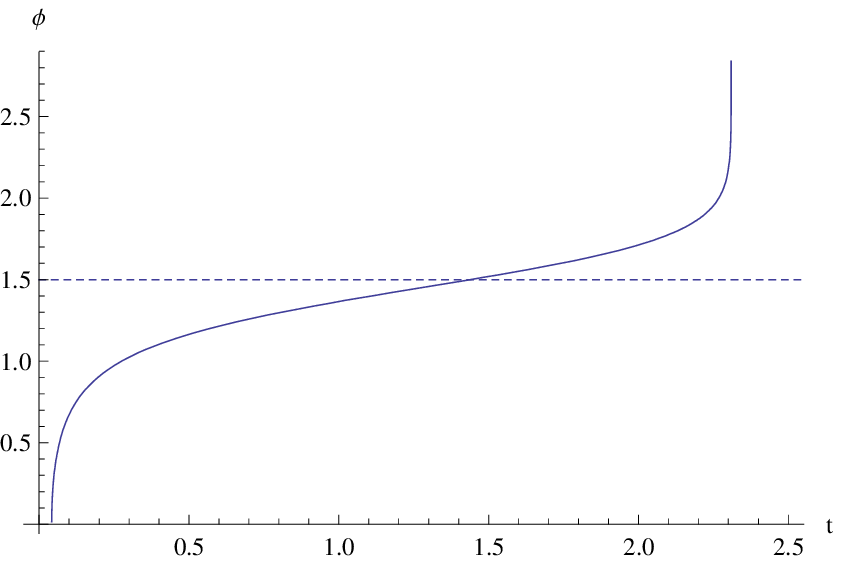,width=7cm}
\hspace{1cm} \epsfig{figure=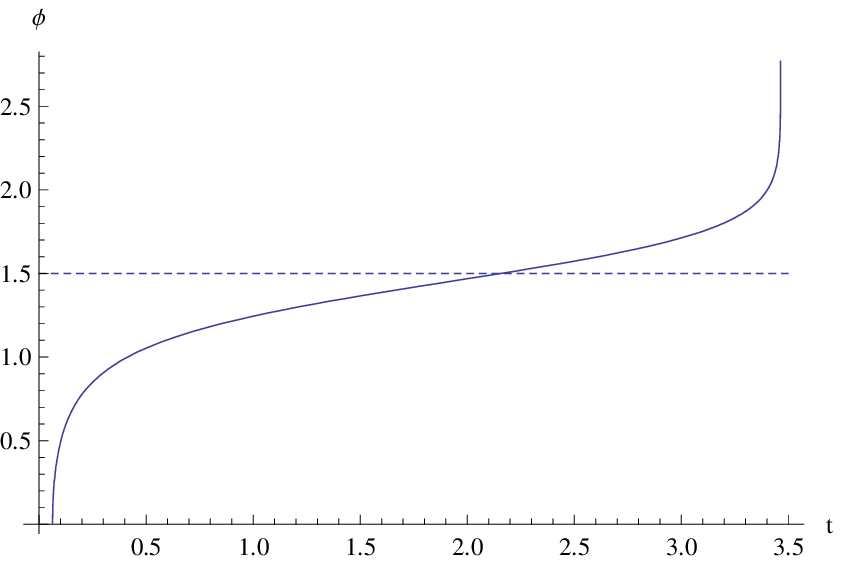,width=7cm}
\end{tabular}
\caption{\footnotesize  Up: the figures show the classical scale factor (solid line), the expectation value of the scale factor (small-dashed line) and the Bohmian scale factor (large-dashed line). Down: the classical scalar field (solid line) and the expectation value or its Bohmian version (dashed line) versus time. The figures are plotted for the numerical values $\gamma = 6$, $P_0= 1$, $C=2$, $a_0=1$, and we set the equation of state parameters $\alpha=-1$ (left) and $\alpha=-1/3$ (right). After examining some other values for this parameter, we verify that the general behavior of the curves is repeated.}\label{fig2}
\end{figure}
Figure \ref{fig2} shows the behavior of the classical scale factor (\ref{U}),
the quantum mechanical expectation value of the scale factor
(\ref{AU}) and its Bohmian version (\ref{BH}) versus time for some
typical numerical values of the parameters. Also, the behavior of
the classical scalar field (\ref{W}) and its quantum mechanical
expectation value is shown in this figure. The origin of the
singularity avoidance may be understood by the existence of the
quantum potential which corrects the classical equations of motion.
To get an approximate scheme of this issue let us neglect the
$\phi$-terms in (\ref{BA}) because of the constant value for $\phi$
from (\ref{BI}). Now, inserting the relation (\ref{BH}) in
(\ref{BC}), we can find the quantum potential in terms of the scale
factor as
\begin{equation}\label{CC}
{\cal Q}(a)=\frac{(3\alpha-1)(3\alpha-2)}{48}a^{-3}.\end{equation}
It is obvious from this equation that the quantum effects are
important for small values of the scale factor and in the limit of
the large scale factor can be neglected. Therefore, asymptotically
the classical behavior is recovered. In this sense we can extract a
repulsive force from the quantum potential (\ref{CC}) as
\begin{equation}\label{DD}
F_a=-\frac{\partial {\cal Q}}{\partial
a}=\frac{1}{16}(3\alpha-1)(3\alpha-2)a^{-4},\end{equation} which may
be interpreted as being responsible of the avoidance of singularity.
For small values of $a$ (near the big-bang singularity), this
repulsive force takes a large magnitude and thus prevents the scale
factor (and then the scalar field) to evolve to the classical
singularity $(a\rightarrow 0, \phi\rightarrow \pm \infty)$.
\section{Conclusions}
In this paper we have studied the classical and quantum dynamics of
a scalar-metric cosmological model coupled to a perfect fluid in the
context of the Schutz' representation. The use of the Schutz'
formalism for perfect fluid allowed us to introduce the only
remaining matter degree of freedom as a time parameter in the model.
In terms of this time parameter, we have obtained the corresponding
classical cosmology by evaluating the dynamical behavior of the
cosmic scale factor and the scalar field. We have seen that the
evolution of the universe based on the classical picture represents
a late time power law expansion coming from a big-bang singularity.
We then dealt with the quantization of the model in which we saw
that the classical singular behavior will be modified. In the
quantum model, we showed that the SWD equation can be separated and
its eigenfunctions can be obtained in terms of analytical functions.
By an appropriate superposition of the eigenfunctions, we
constructed the corresponding wave packet. The wave function  in
this case shows a pattern in which there are two possible quantum
states from which our present universe could have evolved and
tunneled in the past from one state to another. The time evolution
of this wave packet represents its motion along the larger
$a$-direction while the scalar field $\phi$ remains almost constant.
As time passes, our results indicated that the wave packets disperse
and the minimum width being attained at $T = 0$, which means that
the quantum effects are important for small enough $T$,
corresponding to small $a$. The avoidance of classical singularities
due to quantum effects, and the recovery of the classical dynamics
of the universe are another important issues of our quantum
presentation of the model. These questions have been investigated by
two different methods. The time evolution of the expectation value
of the dynamical variables and also their Bohmian counterparts have
been evaluated in the spirit of the many-worlds and ontological
interpretation of quantum cosmology respectively. We verified that a
bouncing singularity-free universe is obtained in both cases. The
use of the ontological interpretation has allowed us to understand
the origin of the avoidance of singularity by a repulsive force due
to the existence of the quantum potential. The repulsive nature of
this force prevents the universe to reach the singularity.

\end{document}